\documentclass[apj]{emulateapj}
\usepackage{apjfonts}

\gdef\kms{km\,s$^{-1}$}
\gdef\msun{${\rm M}_{\odot}$}
\gdef\ew{EW$_{{\rm H}\alpha}$}
\lefthead{van Dokkum et al.}
\righthead{First results from 3D-HST}
\slugcomment{Submitted to ApJ Letters}
\begin{document}

\title{First Results from the 3D-HST Survey:
The Striking Diversity of
Massive Galaxies at $z>1$}

\author{Pieter G.\ van Dokkum\altaffilmark{1}, Gabriel
Brammer\altaffilmark{2}, 
Mattia Fumagalli\altaffilmark{3},
Erica Nelson\altaffilmark{1},
Marijn Franx\altaffilmark{3},
Hans-Walter Rix\altaffilmark{4},
Mariska Kriek\altaffilmark{5},
Rosalind E.\ Skelton\altaffilmark{1},
Shannon Patel\altaffilmark{3},
Kasper B.\ Schmidt\altaffilmark{4},
Rachel Bezanson\altaffilmark{1},
Fuyan Bian\altaffilmark{6},
Elisabete da Cunha\altaffilmark{4},
Dawn K.\ Erb\altaffilmark{7},
Xiaohui Fan\altaffilmark{6},
Natascha F\"orster Schreiber\altaffilmark{8},
Garth D.\ Illingworth\altaffilmark{9},
Ivo Labb\'e\altaffilmark{3},
Britt Lundgren\altaffilmark{1},
Dan Magee\altaffilmark{9},
Danilo Marchesini\altaffilmark{10},
Patrick McCarthy\altaffilmark{11},
Adam Muzzin\altaffilmark{1},
Ryan Quadri\altaffilmark{11},
Charles C.\ Steidel\altaffilmark{12},
Tomer Tal\altaffilmark{1},
David Wake\altaffilmark{1},
Katherine E.\ Whitaker\altaffilmark{1},
Anna Williams\altaffilmark{13}
}

\altaffiltext{1}
{Department of Astronomy, Yale University, New Haven, CT 06520, USA}
\altaffiltext{2}
{European Southern Observatory, Alonso de C\'ordova 3107,
Casilla 19001, Vitacura, Santiago, Chile}
\altaffiltext{3}
{Leiden Observatory, Leiden University, Leiden, The Netherlands}
\altaffiltext{4}
{Max Planck Institute for Astronomy (MPIA), K\"onigstuhl 17,
69117, Heidelberg, Germany}
\altaffiltext{5}{Harvard-Smithsonian Center for Astrophysics,
60 Garden Street, Cambridge, MA 02138, USA}
\altaffiltext{6}{Steward Observatory,
University of Arizona, Tucson, AZ 85721, USA}
\altaffiltext{7}
{Department of Physics, University of Wisconsin-Milwaukee, P.O. Box 413,
Milwaukee, WI 53201, USA}
\altaffiltext{8}
{Max-Planck-Institut f\"ur extraterrestrische Physik,
Giessenbachstrasse, D-85748 Garching, Germany}
\altaffiltext{9}
{Astronomy Department, University of California, Santa Cruz, CA 95064, USA}
\altaffiltext{10}
{Physics and Astronomy Department, Tufts University, Robinson Hall,
Room 257, Medford, MA, 02155, USA}
\altaffiltext{11}
{Carnegie Observatories, 813 Santa Barbara Street, Pasadena, CA 91101, USA}
\altaffiltext{12}
{California Institute of Technology, MS 249-17, Pasadena, CA 91125, USA}
\altaffiltext{13}
{Department of Astronomy, University of Wisconsin-Madison, 475
North Charter Street, Madison, WI 53706, USA}

\begin{abstract}

We present first results from the 3D-HST program, a near-IR
spectroscopic survey performed with the Wide Field Camera 3 on the Hubble
Space Telescope.  We have used 3D-HST spectra to
measure redshifts and H$\alpha$ equivalent widths (\ew) for a
complete, stellar mass-limited sample of 34
galaxies at $1<z<1.5$ with $M_{\rm star}>10^{11}$\,\msun\
in the COSMOS, GOODS, and
AEGIS fields.
We find that a substantial fraction of massive galaxies at this epoch
are forming stars at a high rate:
the fraction of galaxies with \ew\,$>10$\,\AA\ is 59\,\%, compared to 10\,\%
among SDSS galaxies of similar masses at $z=0.1$.
Galaxies with weak H$\alpha$ emission
show absorption lines typical of $2-4$ Gyr old stellar populations.
The structural parameters
of the galaxies, derived from the associated WFC3 F140W
imaging data, correlate with the presence of H$\alpha$;
quiescent galaxies are compact with high
Sersic index and high inferred velocity
dispersion, whereas star-forming
galaxies are typically large
two-armed spiral galaxies, with low Sersic index.
Some of these star forming galaxies might be progenitors
of the most massive S0 and Sa galaxies.
Our results challenge the idea that galaxies
at fixed mass form a homogeneous population with small scatter in
their properties. Instead we find
that massive galaxies form a highly
diverse population at $z>1$, in marked contrast to the local Universe.

\end{abstract}


\section{Introduction}

In the nearby Universe galaxies with stellar
masses $>10^{11}$\,\msun\ form a homogeneous
population, with small scatter in their properties at fixed mass
(e.g., {Djorgovski} \& {Davis} 1987; {Blanton} {et~al.} 2003; {Kauffmann} {et~al.} 2003a).
This homogeneity is somewhat
puzzling in the context of standard models
of galaxy formation, as in these models
star formation and merging extend
to recent epochs (e.g., {De Lucia} {et~al.} 2006). Models that aim to explain
the tight scaling relations of massive galaxies
usually invoke ``dry'' mergers in combination with
feedback from AGN
to prevent gas cooling and star formation
(e.g., {Croton} {et~al.} 2006).

Many studies have measured the properties of massive
galaxies at earlier cosmic epochs to better constrain when and
how they were formed. Interestingly, it was found that massive galaxies 
have red rest-frame optical colors at least
out to $z\sim 2$ ({Bell} {et~al.} 2004; {Faber} {et~al.} 2007; {Brammer} {et~al.} 2011) and
probably beyond (e.g., {van Dokkum} {et~al.} 2006; {Marchesini} {et~al.} 2010).
Many of these galaxies are compact quiescent galaxies
with little or no ongoing
star formation (e.g., {Daddi} {et~al.} 2005;
{van Dokkum} {et~al.} 2008; Damjanov et al.\ 2009),
but at redshifts $z>1$ a subset of the
population has MIPS 24\,$\mu$m fluxes and
near-IR colors characteristic of dust-obscured star-formation
(e.g., {Papovich} {et~al.} 2006; {Williams} {et~al.} 2010; {Brammer} {et~al.} 2011). These galaxies
mascarade as ``dead'' red sequence galaxies as they have very similar
optical colors but in fact have very high inferred star formation
rates.

These apparently star-forming, massive galaxies
are progenitors of at least a subset of massive galaxies
today. In the nearby
Universe strongly star forming
galaxies are typically gas-rich mergers, but several
studies have argued that at higher redshift such galaxies 
more resemble ``scaled-up'' spiral galaxies than local mergers
({Wolf} {et~al.} 2005; {Muzzin} {et~al.} 2010; {Elbaz} {et~al.} 2011).
Interpreting these galaxies, and their
relation to the compact quiescent galaxies which exist at the same
epoch, has been hampered by a lack of mass-complete samples with
homogeneous data at the redshifts of interest. Most existing samples
are luminosity- rather than mass-selected,
are based on photometric redshifts, lack rest-frame optical
morphological information, and/or lack well-calibrated
star formation diagnostics.

In this Letter, we construct and study
a spectroscopic stellar mass-limited sample of
galaxies at $1<z<1.5$, in order to quantify the properties of
massive galaxies when the Universe was $\approx 5$\,Gyr old.
We use data from the
3D-HST survey (GO-12177, GO-12328),
an HST/WFC3 Treasury program that provides
rest-frame optical spectroscopy and imaging at $0\farcs 13$
resolution for thousands of distant galaxies.


\section{Data and Analysis}

The data presented here were obtained in the context of 3D-HST, a 248-orbit
HST Treasury program in Cycles 18 and 19. 3D-HST is a wide-field,
two-orbit depth survey of four (GOODS-South, UDS, AEGIS,
and COSMOS) of the CANDELS fields ({Grogin} {et~al.} 2011, {Koekemoer}
{et al.} 2011) with the Wide Field
Camera 3 (WFC3) G141 grism. The fifth CANDELS field, GOODS-North,
was observed by
the Cycle 17 program GO-11600 (PI: Weiner);
these data are included in the 3D-HST
project. The grism provides spatially-resolved
spectra of all objects in the
WFC3 field, covering the wavelength range 1.15\,$\mu$m -- 1.65\,$\mu$m 
with point-source spectral resolution $R\approx 130$. 
Accompanying
direct imaging in the F140W filter provides the necessary information for
wavelength calibration, as well as photometric and morphological information.
A typical two-orbit pointing 
comprises $\approx 5100$\,s of integration time in
the G141 grism and $\approx 800$\,s in the
F140W direct imaging filter. In addition to the WFC3 near-IR spectroscopy
and imaging, 3D-HST
provides parallel ACS G800L grism spectroscopy and accompanying
F814W imaging in the
optical. The survey is described in G.~Brammer et al.,
in preparation.

The F140W data were reduced using standard procedures for
WFC3 imaging data, and resampled to a rectangular output grid with
$0\farcs 06$ pixels. This fine sampling is justified as the
four individual exposures in each visit are shifted
with respect to each other by an integer + 0.5 pixels in $x$ and $y$.
Spectra were extracted from the grism data using the aXe software 
({K{\"u}mmel} {et~al.} 2009), in a
similar way as described by {van Dokkum} \& {Brammer} (2010) and {Atek} {et~al.} (2010).
Details on the specific 3D-HST reduction
will be provided in G.~Brammer et al., in preparation.
The spectra were combined with
photometry at other wavelengths, using publicly available photometric
catalogs of {Whitaker} {et~al.} (2011) (COSMOS, AEGIS),
{Kajisawa} {et~al.} (2011) (GOODS-North), and
{Wuyts} {et~al.} (2008) (GOODS-South).
Next, redshifts and emission line fluxes were determined from
the combined photometric and spectroscopic data, using
a modified version of the EAZY code ({Brammer}, {van Dokkum}, \& {Coppi} 2008).
The slitless
spectra require that the template fitting explicitly takes
the morphologies of the galaxies into account; therefore,
the models are convolved with the F140W direct image averaged
in the spatial direction.
In our fitting procedure emission
lines are ``automatically'' corrected for underlying
stellar absorption.  
Due to the nature of slitless
grism spectroscopy, many spectra are contaminated
to some extent by overlapping
spectra of neighbouring objects. The aXe package provides a quantitative
estimate of the contamination as a function of wavelength,
which can be subtracted from the spectra.

Stellar masses were determined using the FAST code ({Kriek} {et~al.} 2009a),
using
{Bruzual} \& {Charlot} (2003) models and assuming a {Chabrier} (2003) stellar
initial mass function (IMF). We find that the stellar masses
differ by $<0.05$\,dex whether or not the spectra are included
in the fits.
Rest-frame $U-V$ colors were determined directly from the best-fitting spectral
energy distributions, following the procedures outlined in, e.g.,
{Wolf} {et~al.} (2003) and {Brammer} {et~al.} (2011). Effective radii and {Sersic} (1968)
indices were determined from the 
F140W images using GALFIT ({Peng} {et~al.} 2002),
using synthetic PSFs and averaging the measurements from each of
the dithered exposures. The individual measurements are generally
consistent to within $<0.05$\,dex.

\section{A Mass-Limited Spectroscopic Galaxy Sample at $z=1-1.5$}

For this initial paper, we selected galaxies from the 3D-HST survey with
stellar masses $M_{\rm star}>10^{11}$\,\msun, grism redshifts $1<z<1.5$,
and low contamination (less than 20\,\% of the measured flux).
The only other spectroscopic sample of this kind
is the Gemini Deep Deep Survey, which used optical spectroscopy
(Abraham et al.\ 2004).
The upper limit on the redshift selection is the highest
redshift for which H$\alpha$ falls in the G141 spectral
range.\footnote{At the resolution of the 3D-HST spectra H$\alpha$ and
[N\,{\sc ii}]\,$\lambda{}6583$
cannot be distinguished. Throughout the text ``H$\alpha$''
refers to the combination of these two lines.}
Besides H$\alpha$, strong Mg, Na, and TiO
absorption features fall in the observed wavelength range, allowing
us to measure accurate redshifts for both star-forming and quiescent
galaxies.
From 68 available HST pointings (approximately 50\,\% of the full survey)
this selection gives a sample of 34 galaxies. We note
that our sample is biased
against close pairs, mostly due to the strict contamination criterion.
Fifteen of these objects have a spectroscopic redshift previously
measured with a ground-based telescope. 
There is good agreement between the grism redshifts and
the spectroscopic redshifts: the $1\sigma$ scatter is
$0.004\times (1+z)$.

\begin{figure}[t]
\epsfxsize=8.5cm
\epsffile{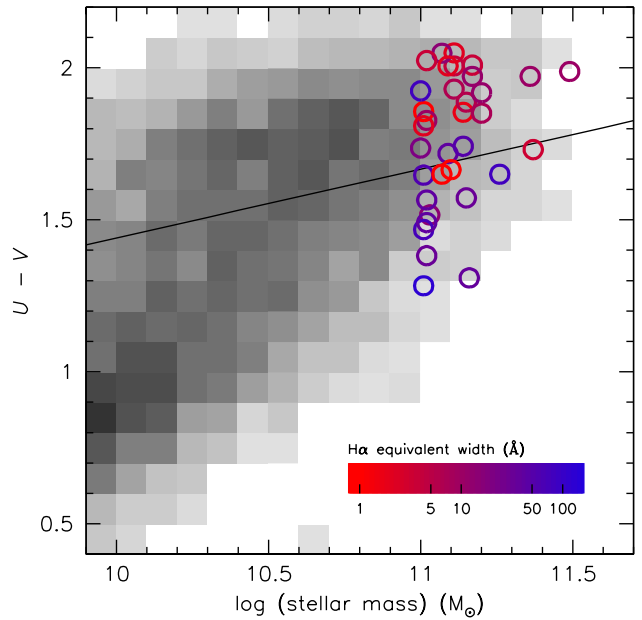}
\caption{\small Color-mass relation of galaxies at $1<z<1.5$
in the NEWFIRM Medium Band Survey (greyscale). The 3D-HST galaxies
are overplotted, color-coded by their H$\alpha$ equivalent width.
The 3D-HST galaxies span a similar range in color as the much
larger NMBS sample with $M>10^{11}$\,\msun.
Star-forming
galaxies are typically bluer than galaxies
with weak H$\alpha$. The line
is the separation between blue and red galaxies from Borch et al.\
(2006), extrapolated to $z=1.25$.
\label{nmbs.plot}}
\end{figure}

In Fig.\ \ref{nmbs.plot} the rest-frame
$U-V$ colors of the galaxies in our sample are compared to those
of $1<z<1.5$ galaxies over a large mass range in
the NEWFIRM Medium Band Survey (NMBS; {Whitaker} {et~al.} 2011).
The 3D-HST galaxies span a similar range in rest-frame
$U-V$ color as massive galaxies in the NMBS. This  range is
much smaller at $M>10^{11}$\,\msun\ than at
lower masses: all massive galaxies
are red compared to blue cloud galaxies at $M\sim 10^{10}$\,\msun,
which have $U-V \sim 0.8$. 
\noindent
\noindent
\begin{figure*}[p]
\epsfxsize=17.5cm
\epsffile{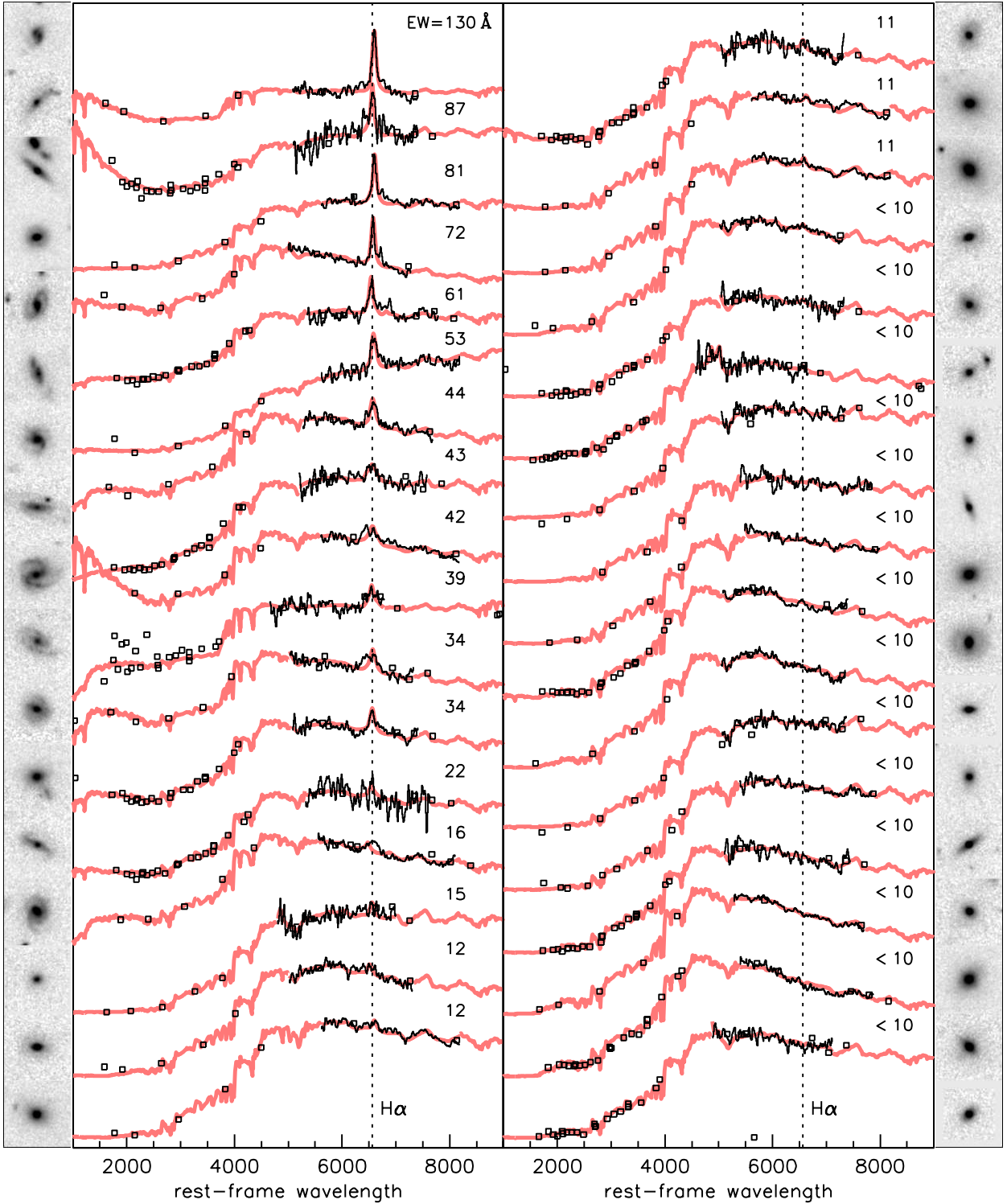}
\caption{\small HST/WFC3 F140W images and G141 grism spectra
of galaxies
with $M_{\rm star}>10^{11}$\,\msun\ and $1<z<1.5$.
The spectra were averaged
in the spatial direction to optimize the S/N ratio, and smoothed by
a boxcar filter for presentation purposes. The $y$-axis is in
units of $F_{\lambda}$; each spectrum was normalized and offset
with respect to the others. Squares indicate broad- and medium-band
photometry from public catalogs (see text). The orange spectra are
best-fitting EAZY models (Brammer et al.\ 2008) for the continuum
emission and for H$\alpha$. In the grism spectral range the models
were convolved with the morphologies of the galaxies to properly model
the spectral resolution.
The galaxies are ordered by decreasing H$\alpha$ equivalent width. 
The spectra are of high quality. The fraction of galaxies with
strong H$\alpha$ emission is much higher than in the
nearby Universe. Galaxies with strong H$\alpha$ emission
are often two-armed spiral galaxies. Galaxies with weak or undetected
H$\alpha$ typically have an early-type (E, S0, or Sa) morphology.
\label{sample.plot}}
\end{figure*}
\section{Spectral Features}

The WFC3/G141 spectra and WFC3/F140W images of the galaxies are shown
in Fig.\ \ref{sample.plot}, ordered by decreasing rest-frame
H$\alpha$ equivalent width.
We detect H$\alpha$ emission in 20 galaxies, or 59\,\% of the sample.
The rest-frame
equivalent widths for the
detected galaxies range from 10\,\AA\ to 130\,\AA.
The immediate implication
is that a substantial fraction of massive galaxies at $1<z<1.5$ is
likely forming stars.
The H$\alpha$ lines in nearby galaxies in this mass range are
substantially weaker: the fraction of galaxies with \ew\,$>10$\,\AA\
above the same mass limit at $z=0.1$ is only $\approx 10$\,\%
({Tremonti} {et~al.} 2004).
Note that the H$\alpha$ lines are typically broad in the 3D-HST
spectra. This
is not due to velocity broadening but rather
to ``morphological broadening'':
in slitless spectroscopy spectral features are effectively
images of the galaxy in that particular wavelength. This will be discussed
further in \S\,\ref{disc.sec}.

\begin{figure}[htbp]
\epsfxsize=8.5cm
\epsffile{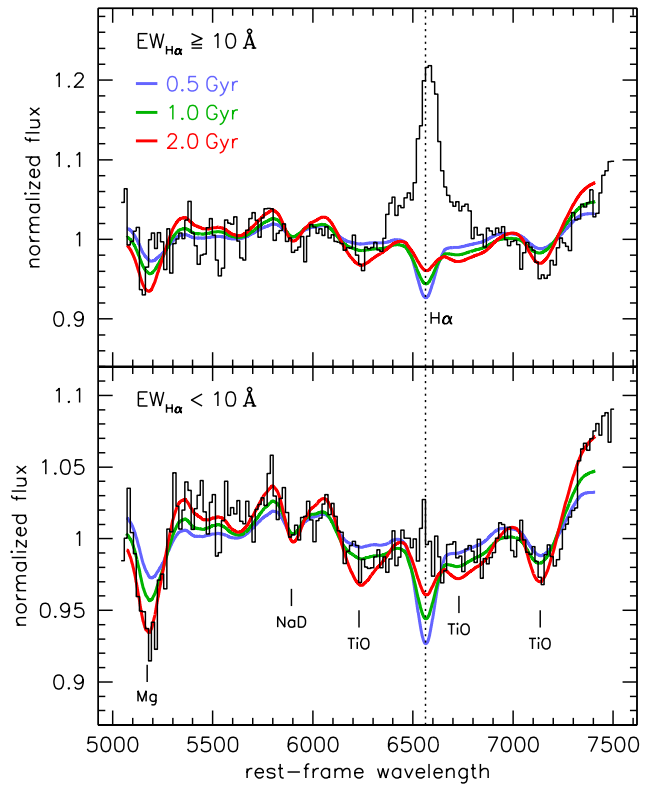}
\caption{\small Stacked spectra of the 20
massive galaxies at $1<z<1.5$ with \ew\,$\geq$\,10\,\AA\
(top panel) and of the 14 galaxies with \ew\,$<$\,10\,\AA\ (bottom panel).
The spectra were normalized by dividing them
by a second-order polynomial fit.
Colored lines are stellar population synthesis models with
different ages (see text), 
smoothed to the resolution of
the stacked spectra. The strong stellar absorption
features of the galaxies with weak H$\alpha$
require luminosity-weighted ages of $2-4$\,Gyr.
\label{stack.plot}}
\end{figure}

The galaxies with weak H$\alpha$ emission have strong stellar absorption
features typical of intermediate age stellar
populations. This is demonstrated in Fig.\
\ref{stack.plot}, which compares the averaged
rest-frame spectra of galaxies with strong and weak H$\alpha$.
The stellar absorption features Mg$_b$, Na\,D, and several TiO bands
are clearly detected in the stacked spectrum of the weak H$\alpha$
galaxies -- for the first time at these redshifts. 
The colored lines in Fig.\ \ref{stack.plot}
are stellar population synthesis models of {Vazdekis} {et~al.} (2010)
with metallicity [M/Z]\,=\,0.22 and different ages. The
best-fitting age for the galaxies with low star formation
is $2$\,Gyr;
models with ages of up to 4\,Gyr also provide good fits. We note
that this is an average and that the galaxies probably have a range
of ages.
%
\noindent
\noindent
\begin{figure*}[htb]
\epsfxsize=14cm
\epsffile{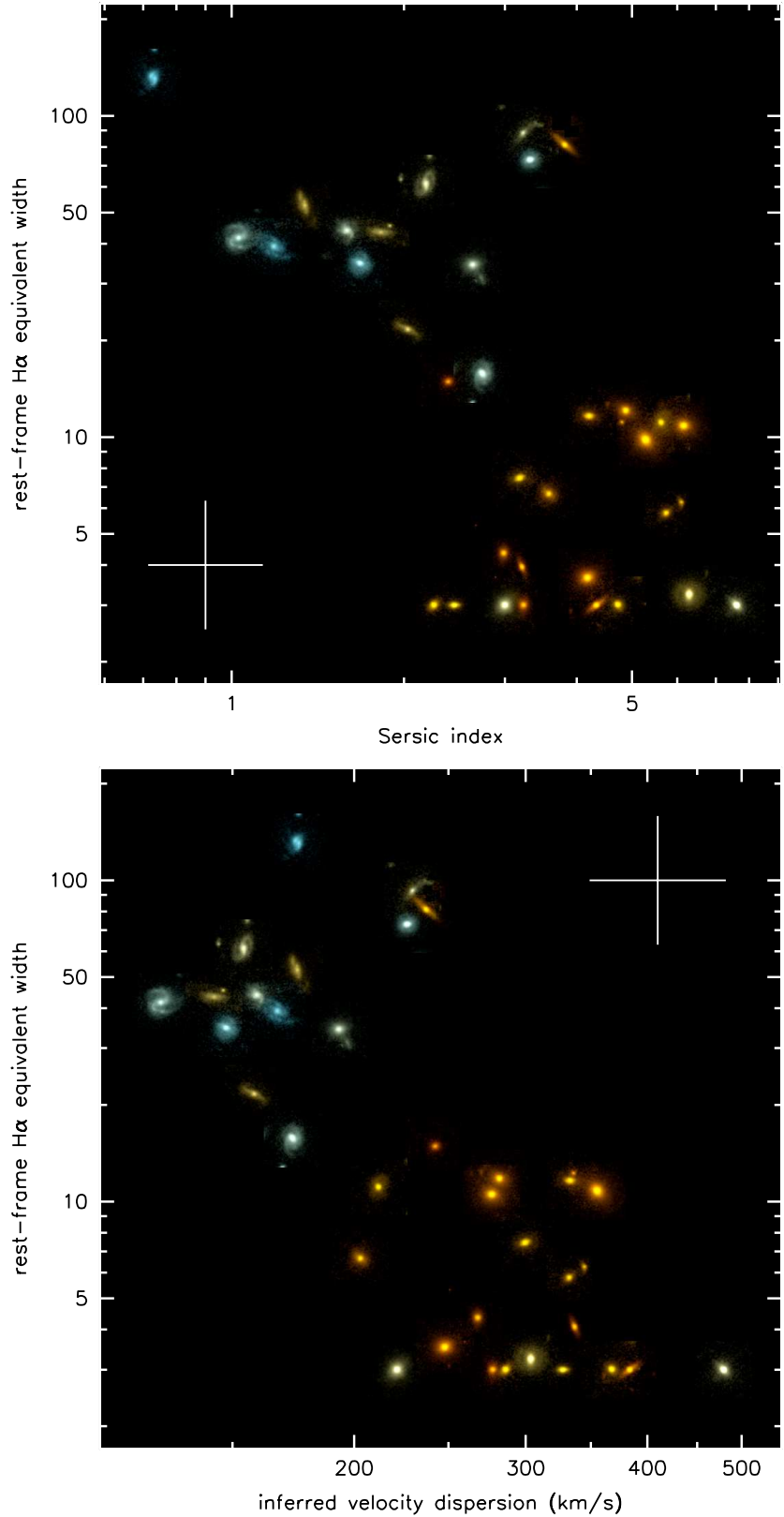}
\caption{\small 
{\em Top panel:} Relation
between \ew, Sersic index, and morphology.
The F140W images are color-coded according to
their average $U-V$ color. 
{\em Bottom panel:} Relation between \ew, velocity dispersion, and
morphology. The velocity dispersions were inferred from the
stellar masses, effective radii, and Sersic $n$ values of the
galaxies. The errorbars denote the typical uncertainties, as
determined from a combination of formal errors and varying the
fitting methodology.
Sersic index, velocity dispersion, and morphology all
correlate with the strength of H$\alpha$. All these parameters
span large ranges: massive galaxies exhibit a
striking
diversity at $1<z<1.5$, ranging from large star-forming spiral galaxies
to compact quiescent galaxies. 
\label{ewn.plot}}
\end{figure*}

As can be seen in Fig.\ \ref{nmbs.plot}
the galaxies with strong H$\alpha$ emission are slightly
bluer (by $\approx 0.3$\,mag in $U-V$)
than the galaxies with weak H$\alpha$. 
The fact that the color difference is
relatively small is because the star-forming galaxies
have more dust: gaxies with \ew\,$<10$\,\AA\ have
a mean best-fit $\langle A_V\rangle
= 0.5 \pm 0.1$ whereas galaxies with \ew\,$>10$\,\AA\
have $\langle A_V\rangle = 1.0 \pm 0.2$
(see also, e.g., {Labb{\'e}} {et~al.} 2005; {Papovich} {et~al.} 2006; {Brammer} {et~al.} 2011).

\section{Star Formation, Structure, and the diversity
of massive galaxies}

Besides the number of objects with
strong H$\alpha$ emission,
a striking aspect of Fig.\ \ref{sample.plot}
is that the morphologies of the galaxies
correlate with the H$\alpha$ line strength.
Massive $1<z<1.5$
galaxies with the highest star formation
rates tend to be ``grand design'' spiral galaxies, whereas those
with no detected H$\alpha$ emission tend to be early-type galaxies.
This result is qualitatively similar to trends at $z=0$, although
massive galaxies with \ew\,$>10$\,\AA\ are
rare in the nearby Universe (see \S\,3).

We quantify this correlation between galaxy structure and star formation
rate in the top panel of
Fig.\ \ref{ewn.plot}, which shows the relation between
\ew\ and {Sersic} (1968) index.
The Sersic index is a quantitative measure of galaxy structure,
and is a proxy for the bulge-to-disk ratio: galaxies
dominated by disks have $n\sim 1$ and galaxies dominated by bulges
have $n\sim 4$. Splitting the galaxies in two equal bins,
the median Sersic index of the 17 galaxies with \ew\,$>12$\,\AA\ is 2.2,
whereas it is 4.2 for the 17 galaxies with \ew\,$<12$\,\AA. According to
the Mann-Whitney test this difference
is significant at the $>99$\,\% confidence level.
We infer that star formation in massive galaxies at $1<z<1.5$ typically
takes place in disk-dominated (spiral) galaxies.
This result is consistent with
recent suggestions that many star-forming galaxies at $z>1$ are ``scaled-up''
versions of nearby galaxies (e.g., {Elbaz} {et~al.} 2011).

The bottom panel of Fig.\ \ref{ewn.plot} shows the relation between
\ew\ and the inferred velocity dispersions of the galaxies,
which are derived from their
stellar masses, effective radii, and Sersic indices following
{Bezanson} {et~al.} (2011). The inferred
dispersions are a measure of the compactness of the stellar
components of galaxies as they are proportional
to $\sqrt{M/r_e}$ (with a Sersic-dependent correction factor).
Again, there is a clear relation between these quantities:
the median inferred dispersion of the half of the sample with
the strongest H$\alpha$ is
$174^{+30}_{-25}$\,km/s whereas it is
$299^{+51}_{-43}$\,km/s for the galaxies with the weakest
H$\alpha$.
The relation between \ew\ and dispersion
has much smaller scatter than the relations of \ew\ with
effective radius and mass separately.
The trend in Fig.\ \ref{ewn.plot}
is qualitatively consistent with the relation between
estimated specific star formation rate (SSFR) and inferred dispersion
found by {Franx} {et~al.} (2008).

Figure \ref{ewn.plot} showcases the large variety of
galaxies. The range in \ew\
among massive galaxies with {\em detected} H$\alpha$
is a factor of 12, and it is obviously even larger when undetected
galaxies are taken into account (see Fig.\ \ref{stack.plot}).
The structure and
morphologies of galaxies also show a large range, going from large
spiral galaxies with high \ew\ to compact early-type galaxies with
low \ew. The Sersic indices span the full range between
disk- to bulge-dominated values ($n\sim 1$ to $n\sim 4$) and
the inferred dispersions range from $\lesssim 150$\,\kms to $\gtrsim
400$\,\kms.

\section{Discussion}
\label{disc.sec}

We have shown that the 3D-HST Treasury program is providing
high quality near-IR
spectroscopy and accompanying imaging for galaxies in the
crucial epoch $1<z<3$, when the cosmic star formation rate peaked
(e.g., {Bouwens} {et~al.} 2007). In this first paper we focused on a
stellar mass-limited sample of 34 galaxies at $z=1-1.5$, to study the
properties of massive
galaxies when the Universe was 4--6\,Gyr old. This is the first
study of the H$\alpha$ emission and rest-frame optical
morphologies of a complete, mass-limited galaxy
sample at $z>1$.

The most striking result of our study is the
diversity of the spectra and structure of massive galaxies
at $z>1$: a large fraction has strong H$\alpha$ emission, whereas
others have absorption features characteristic of relatively
old stellar populations. Similarly, the morphologies,
Sersic indices, and implied velocity dispersions show a large range.
These results are broadly consistent with
previous studies that were based on photometric redshifts,
the SED shapes of galaxies,
and/or imaging of lower quality
(e.g., {Franx} {et~al.} 2008; {Williams} {et~al.} 2010; {Wuyts} {et~al.} 2011b;
Weinzirl {et~al.} 2011). 
The simplest interpretation is that at $z>1$ we are
entering the epoch when
massive galaxies were undergoing rapid evolution. Specifically,
the star-forming disks may be progenitors of some
of today's most massive S0 and Sa galaxies.

{Noeske} {et~al.} (2007) suggested that the star formation rates of
galaxies are tightly coupled to their mass and redshift, and
it is interesting to compare the range in \ew\ in our sample
to the scatter in their ``star formation main sequence''.
At fixed mass {Noeske} {et~al.} (2007) find 
a 68\,\% range in star formation rates
of a factor of $\sim 4$ (among galaxies
with clear signs of star formation); if
we limit the sample to the central 68\,\% of the
distribution of galaxies with \ew\,$>10$\,\AA\
we find a larger range of a factor of six. More to the point, within our
sample
the scatter in \ew\ can be significantly reduced by
considering the structure of the galaxies. The 68\,\%
range in \ew\ among the 11 galaxies with inferred $\sigma<200$\,\kms\ is
only a factor of 1.7. We therefore follow earlier work in suggesting
that velocity dispersion (or surface density) is a more fundamental
parameter than mass in determining the properties of galaxies
(see, e.g., {Kauffmann} {et~al.} 2003b; {Franx} {et~al.} 2008; {Bezanson} {et~al.} 2011).
\noindent
\begin{figure}[b]
\epsfxsize=8cm
\epsffile{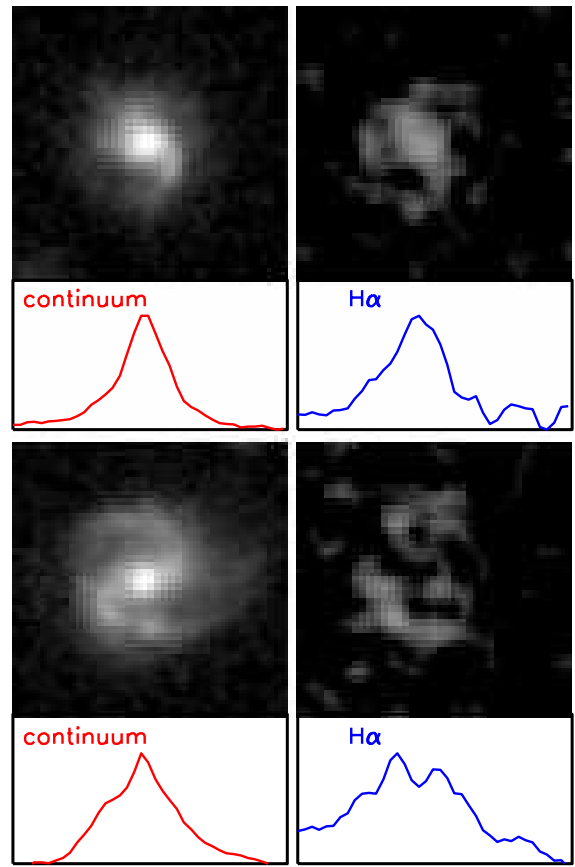}
\caption{\small Spatial distribution of
continuum emission in the F140W filter compared to the spatial
distribution of the H$\alpha$ line emitting gas. The H$\alpha$
images were made by subtracting a polynomial fit from the
two-dimensional G141 grism spectra. The boxes are $4\farcs 7
\times 4\farcs 7$, or 39\,kpc\,$\times$\,39\,kpc. The line profiles
show the flux distributions averaged along the $y$-axis.
The line emission in these particular galaxies is
extended, and the galaxies are building up their disks.
We also find galaxies with more compact H$\alpha$ distributions.
\label{spiral.plot}}
\end{figure}

The growth rate of the star-forming galaxies is substantial; using
standard prescriptions to correct for extinction toward H\,{\sc ii}
regions ({Calzetti} {et~al.} 2000; {Wuyts} {et~al.} 2011a), we find a median stellar
mass increase due to star formation
of $\sim 50$\,\% per Gyr for the galaxies with \ew\,$>10$\,\AA.
An important question is {\em where} in the galaxies the star formation
is occurring, that is, which structural component of massive galaxies
is in the process of formation at $1<z<1.5$. Due to the nature
of grism spectroscopy the 
3D-HST data provide 2D emission line maps at the spatial resolution of
HST. Two examples are shown in Fig.\ \ref{spiral.plot}: in these galaxies
the star formation appears to trace the spiral arms, similar to
spiral galaxies in the nearby Universe.
A quantitative analysis of the spatial extent of the emission
line gas is beyond the scope of this Letter, but we note here that
in cases such as those shown in Fig.\ \ref{spiral.plot}
the spatial extent of the H$\alpha$ emission rules out
dominant contributions from
active nuclei to the integrated line fluxes.

A key open question is what  drives the diversity of
massive galaxies at $z>1$. At fixed stellar mass, we see large, star
forming spiral galaxies and very compact galaxies in which star formation
has apparently ceased. If AGN feedback is responsible for shutting off
star formation in massive galaxies it is clearly more effective
in some galaxies than in others. It may be that AGN feedback
correlates with black hole mass, which
correlates better with velocity dispersion than with stellar mass
(e.g., {Magorrian} {et~al.} 1998). It will
also be interesting to study correlations with other parameters, such
as the environment, at fixed stellar mass and at fixed (inferred)
velocity dispersion. Finally, it will be important to extend
this study to lower masses and to higher redshifts. Star forming
galaxies that have been studied
at $z\gtrsim 2$ tend to have higher \ew\ and
also more irregular morphologies than the galaxies studied here
(e.g., {Erb} {et~al.} 2006; {Kriek} {et~al.} 2009b;
{F{\"o}rster Schreiber} {et~al.} 2011), and it will be interesting to see
whether a broader selection of galaxies would only increase the
dynamic range in Fig.\ \ref{ewn.plot} or (also) increase the
scatter.

Such studies will be possible with the full 3D-HST dataset, which
will provide WFC3 and ACS spectroscopy and imaging of large
samples of
galaxies at $z>1$. Combined with deep imaging from the CANDELS
project and the wide array of ancillary data in the survey fields,
3D-HST will provide a qualitatively new way of surveying the
heyday of galaxy formation at $1<z<3$.



\end{document}